# Symmetry invariance for adapting biological systems

Oren Shoval[1], Uri Alon[1], and Eduardo Sontag[2]

November 3, 2018

## 1   Introduction

We study in this paper certain properties of the responses of dynamical systems to external inputs. Our results are purely mathematical and thus of wide applicability, but our motivation arises from molecular systems biology. Indeed, the behavior, adaptability, and survival of organisms depends critically upon their capability to formulate appropriate responses to chemical and physical environmental cues. In particular, signal transduction and gene regulatory networks in individual cells mediate the processing of measured chemical concentrations and physical conditions, such as ligand concentrations or stresses, eventually leading to regulatory changes in metabolism and gene expression. Often, the ultimate goal of these changes is to maintain a narrow range of concentration levels of vital quantities (homeostasis, adaptation) while at the same time appropriately reacting to changes in the environment (signal detection). Much theoretical, modeling, and analysis effort has been devoted to the understanding of these questions, traditionally in the context of steady-state responses to constant or step-changing stimuli.

In this work, we are concerned with questions that complement the analysis of simple temporal inputs and steady-state responses, focusing on certain properties of *transient* behaviors, both for simple stimuli like step changes and for more complex time-varying input profiles. The study of transient responses is of a central concern in cell biology, since behavior at the time-scale of signaling may have important consequences for cell survival. Moreover, typical signals encountered by cells in their natural environments may well exhibit interesting temporal information, and thus characterizing responses to fluctuating temporal patterns may provide new insights regarding cell behavior. Such responses and the analysis of transient behavior, can also help rule out postulated mechanisms when tested through modern experimental tools which allow for fine spatiotemporal resolution in measurements. This type of model falsification is of course routine, but the broader framework allows for testing of a richer class of "phenotypes" and may thus help guide searches for yet-unknown molecular mechanisms.

The immediate motivation for this work is the recent discovery of an important transient property, related to Weber's law in psychophysics: *fold-change detection (FCD)* in adapting systems, the property that scale uncertainty does not affect responses. FCD appears to play an important role in key signaling transduction mechanisms in eukaryotes, including the ERK and Wnt pathways, as well as in *Escherichia ecoli* and possibly other prokaryotic chemotaxis pathways [1–3]. The mathematical analysis of FCD was started in [3,4]. In this paper, we provide further theoretical results regarding this property. Far more generally, we develop a necessary and sufficient characterization of adapting systems whose transient behaviors are *invariant under the action of a set (often, a group) of symmetries in their sensory field*. A particular instance is FCD, which amounts to invariance under the action of the multiplicative group of positive real numbers. Our main result is framed in terms of a notion which considerably extends equivariant actions of compact Lie groups.

---

1 Department of Molecular Cell Biology, Weizmann Institute of Science, Rehovot, Israel
2 Department of Mathematics, Rutgers University, Piscataway, NJ, USA

This paper is organized as follows. Section 2 introduces the main definitions and the statement of the main result of this paper. Section 3 explains illustrates how the main result can be used to check invariance in a number of simple examples. Section 4 has the proof of the main theorem, as well as a self-contained review of some key concepts needed from nonlinear control theory. Section 5 fills-in a stability proof that is needed in order to justify that several of our simple examples are indeed adapting systems. Section 6 compares feedforward and feedback architectures, in the context of the "internal model principle" of control theory. Section 7 provides a simple result showing that search strategies in sensory fields subject to symmetries are invariant, provided that the underlying system itself be invariant.

## 2   Notations, definitions, and statement of main theorem

We study dynamical systems with inputs and outputs:

$$\dot{z} = F(z, u), \qquad y = h(z), \tag{1}$$

where $F$, $h$ are functions which describe respectively the dynamics and the read-out map. Equation (1) is meant as shorthand for

$$\frac{dz}{dt}(t) = F(z(t), u(t)), \qquad y(t) = h(z(t)).$$

Here, $u = u(t)$ is a generally time-dependent input (also called, depending on the context, a "stimulus" or "excitation") function, $z(t)$ is an $n$-dimensional vector of state variables, and $y(t)$ is the output ("response" or "reporter" variable).

The paradigm of studying systems in the form (1) is standard in in control systems theory [5].

Typically, $y(t)$ is just a read-out of one of the components, $z_i(t)$, of $z(t)$. However, it is also sometimes natural to take a more complicated function $y(t) = h(z(t))$ of the coordinates of $z$ than just picking an individual $z_i$. For instance, suppose that $z_1(t)$ represents the concentration of the free form of an enzyme $E$, that $z_2(t)$ is the concentration of $E$ complexed with some substrate, and that these two species are indistinguishable by a Western blot assay measurement. Then the sum $y = h(z) = z_1 + z_2$ might be the reporter variable of interest. More generally, the theory does not change substantially if we allow the output variable $y(t)$ to be a function of the current input as well as on the current state, $y = h(z, u)$. It is notationally and technically more convenient to take $y = h(z)$, so we do that here. In any event, one could add a new variable $z_{n+1}$ to $z = (z_1, \ldots, z_n)$ and a differential equation $\varepsilon \dot{z}_{n+1} = h(z, u) - z_{n+1}$ which (provided $0 < \varepsilon \ll 1$) quickly equilibrates to $y = z_{n+1} = h(z, u)$, and with this small modification, now $y$ depends only on a single coordinate, $z_{n+1}$, of an extended state vector $(z_1, \ldots, z_{n+1})$, and does not directly depend on the current input.

In order to describe positivity of variables as well and other constraints, we introduce the following additional notations. States, inputs, and outputs are constrained to lie in particular subsets, which we call $\mathbb{Z}$, $\mathbb{U}$, and $\mathbb{Y}$ respectively, of Euclidean spaces $\mathbb{R}^n, \mathbb{R}^m, \mathbb{R}^q$. For example, $\mathbb{U} = \mathbb{R}_{>0}$ means that the input values must be scalar ($m = 1$, $\mathbb{U} \subset \mathbb{R}^1$) and positive. The functions $f, h$ are differentiable. We will assume that for each piecewise-continuous input $u : [0, \infty) \to \mathbb{U}$, and each initial state $\xi \in \mathbb{Z}$, there is a (unique) solution $z : [0, \infty) \to \mathbb{Z}$ of (1) with initial condition $z(0) = \xi$, which we write as

$$\varphi(t, \xi, u),$$

and we denote the corresponding output $y : [0, \infty) \to \mathbb{Y}$, given by $h(\varphi(t, \xi, u))$, as

$$\psi(t, \xi, u)$$

(see [5] for more discussion, properties of ODE's, global existence of solutions, etc.).

Since we are interested in adapting systems, will assume that for each constant input $u(t) \equiv \bar{u}$, there is a unique steady state of the system (which depends on the particular input). We denote this steady state by $\sigma(\bar{u})$. In other words, $\bar{z} = \sigma(\bar{u})$ the unique solution of the algebraic equation $F(\bar{z}, \bar{u}) = 0$. Finally, we will assume that this steady state is *globally asymptotically stable (GAS)*. This means that $\sigma(\bar{u})$ is Lyapunov stable and globally attracting for the system when the input is $u(t) \equiv \bar{u}$:

$$\lim_{t \to \infty} \varphi(t, \xi, u) = \sigma(\bar{u})$$

for every initial condition $\xi \in \mathbb{Z}$. Multi-stable systems may be considered as well, of course, but the definitions to follow become very cumbersome.

We will illustrate our results using the two sets of examples that are shown in Figs. 1 and 2. In

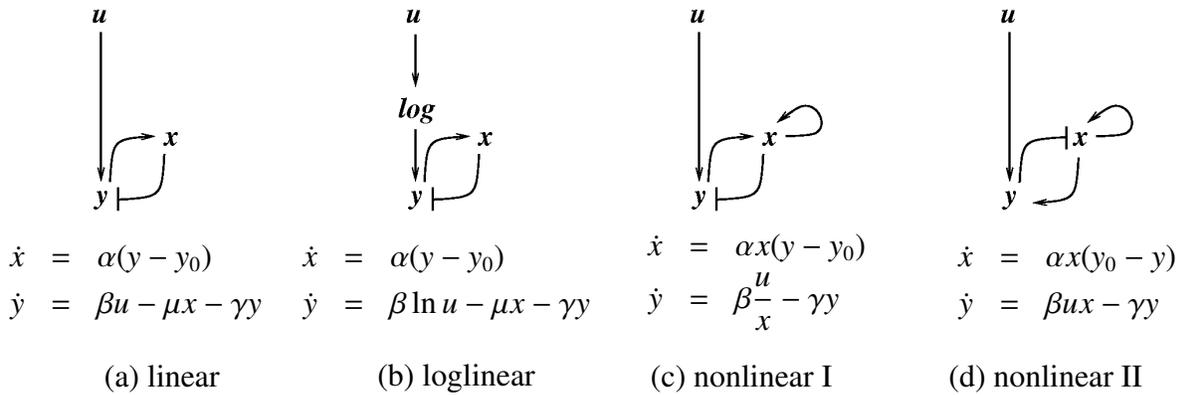

| | | | |
|---|---|---|---|
| $\dot{x} = \alpha(y - y_0)$ | $\dot{x} = \alpha(y - y_0)$ | $\dot{x} = \alpha x(y - y_0)$ | $\dot{x} = \alpha x(y_0 - y)$ |
| $\dot{y} = \beta u - \mu x - \gamma y$ | $\dot{y} = \beta \ln u - \mu x - \gamma y$ | $\dot{y} = \beta \dfrac{u}{x} - \gamma y$ | $\dot{y} = \beta u x - \gamma y$ |
| (a) linear | (b) loglinear | (c) nonlinear I | (d) nonlinear II |

Figure 1: Integral feedback systems (assuming $u > 0$ in (b), and $u, x > 0$ in (c,d))

the equations shown in these figures, the vector $z = (z_1, z_2) = (x, y)$ has dimension $n = 2$ and the output is the second component, $h(z) = y$. In other words, in these examples, $F(z, u) = F(x, y, u)$ is a vector function with two components, which for $n = 2$ we write as $(f(x, y, u), g(x, y, u))$. We have that $f(x, y, u)$ is $\alpha(y - y_0)$ in Fig. 1(a,b), $\alpha x(y - y_0)$ in Fig. 1(c,d), and $\alpha u - \delta x$, in Fig. 2, and $g(x, y, u)$

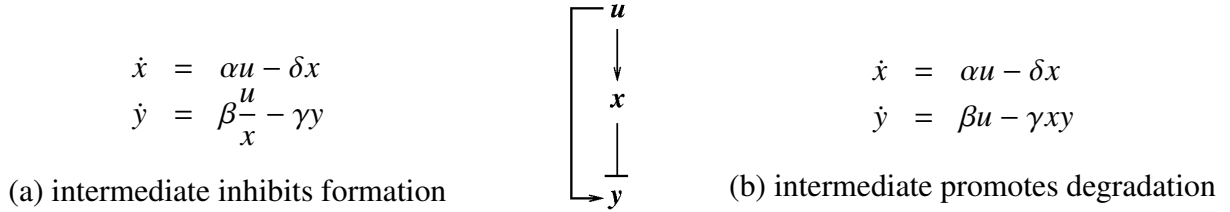

$$\dot{x} = \alpha u - \delta x$$
$$\dot{y} = \beta \frac{u}{x} - \gamma y$$

(a) intermediate inhibits formation

$$\dot{x} = \alpha u - \delta x$$
$$\dot{y} = \beta u - \gamma xy$$

(b) intermediate promotes degradation

Figure 2: Incoherent feedforward loops (IFFL) (assuming $u, x > 0$ in (a))

is $\beta u - \mu x - \gamma y$, in Fig. 1(a), $\beta \ln u - \mu x - \gamma y$, in Fig. 1(b), $\beta \frac{u}{x} - \gamma y$ in Fig. 1(c) and Fig. 2(a), $\beta ux - \gamma y$ in Fig. 1(d), and $\beta u - \gamma xy$ in Fig. 2(b). The constants $\alpha, \beta, \dots$ are positive numbers. *We emphasize that our theory is valid for any dimension n;* these examples are only picked for illustration.

## 2.1 Perfect adaptation

**Definition 2.1** The system (1) *perfectly adapts to constant inputs* provided that the steady-state output $h(\sigma(\bar{u}))$ equals some fixed $y_0 \in \mathbb{Y}$, independently of the particular input value $\bar{u} \in \mathbb{U}$. □

Adaptation (we will just say "adapt" instead of "perfectly adapt to constant inputs" in what follows) means that the steady-state output value is independent of the actual value of the input, if the input is constant. This property may be achieved by differentiating the input signal before it is further processed by the system. However, differentiation is very sensitive to high-frequency noise, and in fact there is no need for differentiation to be explicitly performed: there are several alternative mechanisms, such as those represented in Figs. 1 and 2, integral feedback and incoherent feedforward loops respectively, that are also capable of achieving adaptation.

In integral feedback systems, an internal "memory" variable keeps track of the accumulated (i.e., integrated) difference between the current value $y(t)$ of the response variable and its desired steady-state value $y_0$. A difference, or a nonlinear comparison such as a ratio, is performed between this memory variable and the current input, thus providing an "error" signal that is used to drive the feedback mechanism that brings the system back to its default value. Integral feedback is recognized as a key feature of perfectly adapting biological systems, both at the physiological and cellular level, such as, for example, in blood calcium homeostasis [6], in neuronal control of the prefrontal cortex [7], in the regulation of tryptophan in *E. coli* [8], and in *E. coli* chemotaxis [9].

Fig. 1(a) shows the linear integral feedback configuration (PI, or proportional-integral, control) that is classically treated in control theory. When $u$ is constant, the unique steady state of this system is given by $\bar{x} = (\beta u - \gamma y_0)/\mu$ and $\bar{y} = y_0$. Thus, this system adapts: the steady state value of $y$ is independent of the input $u$. Moreover, since the eigenvalues of any matrix of the form

$$\begin{pmatrix} 0 & \alpha \\ -\mu & -\gamma \end{pmatrix}$$

are negative, it is clear that this system is globally asymptotically stable: $x(t) \to \bar{x}$ and $y(t) \to \bar{y} = y_0$ as $t \to \infty$. (Taking $\alpha$ and $\mu$ both negative instead of positive gives the same conclusions.) Two other integral feedback configurations, also perfectly adapting, are shown in Fig. 1. In (b), a "log-linear system," the only difference with (a) being that the input is logarithmically pre-processed;

this does not change the conclusions of adaptation and stability. In system (c), the memory variable feeds upon itself, and the ratio $u/x$, instead of a difference, is used to compare the current input and memory values. In system (d), the memory variable also feeds upon itself, and the product $ux$ is used in the feedback term to $y$. Both (c) and (d) adapt (to $\bar{y} = y_0$), and $\bar{x} = \beta\mu/(\gamma\bar{y})$ in (c), $\bar{x} = \gamma\bar{y}/(\beta\mu)$ in (d). Stability is a bit more subtle, and is based on a control-Lyapunov approach [5] that recasts (c) as a Hamiltonian system with added damping, see Section 5. (The ratio "$u/x$" in (c) is not a natural choice for biological models; however, one may think of this term as an approximation of a Michaelis-Menten inhibition term $u/(K_m + x)$, with $K_m \ll 1$.)

A different type of architecture is based on *feedforward* as opposed to feedback interconnections. Feedforward circuits are ubiquitous in biology, as emphasized in [10], where they were shown to be over-represented in *E. coli* gene transcription networks, compared to other "motifs" involving three nodes. Similar conclusions apply to certain control mechanisms in mammalian cells [11]. A large number of papers have been devoted to the signal-processing capabilities of the feedforward motif, notably [12] which looked into its properties as a "change detector" (essentially, sensitivity to changes in the magnitude of the input signal), and [13] which studied its optimality with respect to periodic inputs. Comparisons with other "three node" architectures with respect to the trade-off of sensitivity versus noise filtering are given in [14]. Other references on feedforward circuits include [15] (showing their over-representation at the interface of genetic and metabolic networks), [16] (classification of different subtypes of such circuits), and [17] (classification into "time-dependent" versus "dose-dependent" biphasic responses, which are in a sense the opposite of adapted responses).

In particular, in *incoherent feedforward loops (IFFL)*, as in Fig. 2, the input $u$ directly helps promote formation of the reporter $y$ and also acts as a delayed inhibitor, through an intermediate variable $x$. This "incoherent" counterbalance between a positive and a negative effect gives rise, under appropriate conditions, to adaptation. The reference [17] provides a large number of incoherent feedforward input-to-response circuits, which participate in EGF to ERK activation [18,19], glucose to insulin release [20,21], ATP to intracellular calcium release [22,23], nitric oxide to NF-$\kappa$B activation [24], microRNA regulation [25], and many others. Several varieties of IFFL circuits have been often proposed for perfect adaptation to constant signals in biological systems. Notably, the IFFL shown in Fig. 2(b), often called the "sniffer" [26,27], appears in slightly modified forms in models for *Dictyostelium* chemotaxis and neutrophils [28,29], microRNA-mediated loops [30], and *E. coli* carbohydrate uptake via the carbohydrate phosphotransferase system [31] and other metabolic systems [32]. The work [33] shows experimentally and analytically that IFFL's are especially well-suited to controlling protein expression under DNA copy variability. For both systems in Fig. 2, the unique steady state, when the input $u$ is constant, has coordinates $\bar{x} = \alpha u/\delta$ and $\bar{y} = y_0 = \beta\delta/(\alpha\gamma)$. Since $\bar{y}$ is independent of $u$, the system adapts. Global asymptotic stability for (a) follows from the fact that the $x$-subsystem is linear and stable, and the $y$-subsystem is a stable linear system driven by the converging signal $u/x$. For (b) (and several variations of this system), the GAS property is studied in [27].

## 2.2 Invariance

As mentioned in the introduction, we wish to study the invariance of outputs under input transformations. The original motivation was the study of the particular case of scale invariance, which is described through the following thought experiment.

### 2.2.1 A special case: scale-invariance (FCD)

Suppose that a system that adapts has had a chance to "pre-adapt" to a certain constant ("background") level $\bar{u}$ of the input, for $t < 0$, and that now we present the system with the new input $u(t)$, $t \geq 0$. Let $y_1(t)$ be the output function that results. Next imagine that we allowed the same system to pre-adapt, instead, to $p\bar{u}$ for $t < 0$, and we now present the system with $pu(t)$ for $t \geq 0$, where $p$ is some positive scalar. Let $y_2(t)$ be the resulting output. Scale invariance means that the outputs of the two experiments should be the same: $y_1(t) = y_2(t)$ for all $t > 0$. In other words, for any two inputs $u(t)$ and $pu(t)$, as in Fig. 3, and no matter what positive number or "scale" $p$ we picked, the *entire shape of the response*, including amplitude and duration, is identical. As an example, a step change in input from, say, a constant level 2 to a constant level 4, should result in precisely the same output as a step from constant level 5 to constant level 10 (we scaled everything by $p = 2$), Fig. 4(d). On the other hand, a change from, say, level 5 to 25 (which has a fold-change of 5, instead of 2) will typically lead to a different response. Thus, another way to describe this invariance property is to say that the only potentially detectable differences in response are due to fold changes, and this motivates the name *fold-change detection (FCD)* [3, 4]. FCD represents a particular type of adaptation, one in which there is with robustness to scale uncertainty, and it can be found at many levels of biological organization. A weak version is present in the Weber-Fechner law logarithmic sensing feature in psychophysics: many sensory systems (for weight, vision, hearing) produce responses whose maximal amplitude only depends on the *ratio* between the stimulus and a background or starting value, Fig. 4(c). It was also recently discovered that the transient responses of several biological cellular signaling systems [1, 2] display FCD features.

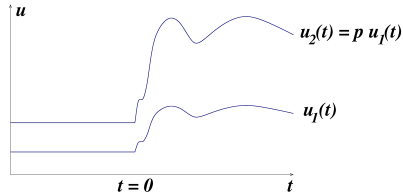

Figure 3: Two scaled inputs

We can formally define the FCD property as follows. Denoting by "$\pi u$" the input $t \mapsto pu(t)$, and by "$\pi\bar{u}$" the constant $p\bar{u}$, the equality of outputs

$$\psi(t, \sigma(\bar{u}), u) = \psi(t, \sigma(\pi\bar{u}), \pi u) \quad \forall\, t \geq 0 \tag{2}$$

should hold for all possible $p > 0$, as well as all constant inputs $\bar{u} \in \mathbb{U}$ and inputs $u = u(t)$ with values in $\mathbb{U}$. (Observe that the requirement that $\bar{u}$ and $u$ take values in $\mathbb{U}$ serves to impose constraints such as, for example, positivity.)

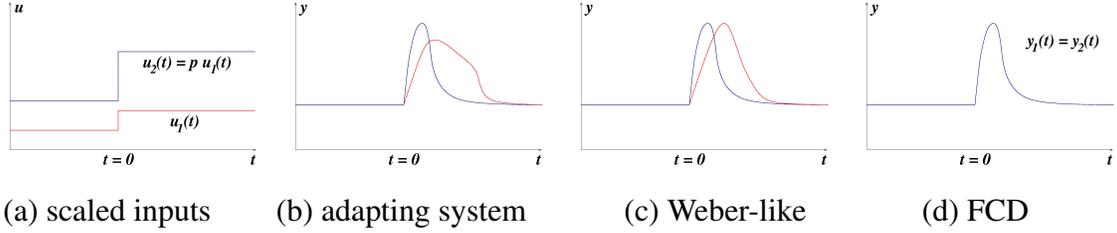

| (a) scaled inputs | (b) adapting system | (c) Weber-like | (d) FCD |

Figure 4: In an adapting system, the response to scaled steps (a) returns to the same adapted value, but may differ in time (b), or may produce the same maximal response (c) or even identical transients (d))

Equipped with this formalism, we may study a far more general question, namely invariance with respect to any set of transformations, not merely scalings but also far more general invariances to sensory field changes: translational, rotational or reflection symmetries, anisotropic dilations, projective transformations, and so forth.

### 2.2.2 General case

Suppose given a set $\mathcal{P}$ of continuous and onto input transformations $\pi : \mathbb{U} \rightarrow \mathbb{U}$:

$$\{\pi : \mathbb{U} \rightarrow \mathbb{U} , \pi \in \mathcal{P}\} .$$

For an input $u(t)$, we denote by "$\pi u$" the function of time that equals $\pi(u(t))$ at time $t$. (For notational simplicity, we write "$\pi u$" rather than "$\pi(u)$" when there is no risk of confusion, but we emphasize that there is no requirement that the mapping $\pi$ be linear.) The continuity assumption is only made in order to insure that $\pi u$ is a piecewise continuous function of time if $u$ is, as technically required when using it as an input to a differential equation. The ontoness assumption, that is, $\pi \mathbb{U} = \mathbb{U}$, is made for technical reasons in the proof of the main theorem.

**Definition 2.2** The system (1) has *response invariance to symmetries in* $\mathcal{P}$ or, for short, is $\mathcal{P}$-*invariant* if (2) holds for all $t > 0$, all inputs $u = u(t)$, all constants $\bar{u}$, and all transformations $\pi \in \mathcal{P}$. □

Under the assumption that that the action of $\mathcal{P}$ is transitive, i.e., for any two $\bar{u}, \bar{v} \in \mathbb{U}$, there is some $\pi$ such that $\bar{v} = \pi \bar{u}$, $\mathcal{P}$-invariance implies perfect adaptation, because the outputs in (2) must coincide at time zero, and any two inputs can be mapped to each other.

**Examples:** In the special case of the transitive action with $\mathcal{P} = \mathbb{U} = \mathbb{R}_{>0}$ and $\pi u = pu$, we obtain the FCD property. Another example of a set of symmetries $\mathcal{P}$ is as follows. Suppose that $\mathbb{U} = \mathbb{R}^n$ and $\mathcal{P}$ consists of all the transformations $\pi_R(u) := Ru$, for $R \in SO(n)$, the special orthogonal group. That is, we transform inputs by multiplication with a rotation matrix $R$. In dimensions 2 or 3, this can be used to impose the requirement that a visual sensing system should react the same if the visual field is rotated. Another slightly different example would be that in which $\mathcal{R} \in O(n)$, the orthogonal group, meaning that we want invariance with respect to reflections as well as rotations.

## 2.3 Equivariances

We now introduce a concept that leads to an effective criterion for checking FCD, and more generally arbitrary $\mathcal{P}$-invariance, *without having to compute the solutions $\psi(t, \sigma(\pi\bar{u}), \pi u)$*. We make the definition for arbitrary systems as in (1), $\dot{z} = F(z, u)$, $y = h(z)$.

**Definition 2.3** Given a system (1) and a set of input transformations $\mathcal{P}$, a parametrized set of differentiable mappings $\{\rho_\pi : \mathbb{Z} \to \mathbb{Z}\}_{\pi \in \mathcal{P}}$ is a $\mathcal{P}$-*equivariance family* provided that, for each $\pi$:

$$F(\rho_\pi(z), \pi u) = \rho'_\pi(z) F(z, u) \quad \text{and} \quad h(\rho_\pi(z)) = h(z) \quad \text{for all } z \in \mathbb{Z}, \, u \in \mathbb{U} \qquad (3)$$

($\rho'_\pi$ = Jacobian matrix of $\rho_\pi$). If (3) holds, we say that the system is $\rho_\pi$-equivariant under the input transformation $\pi$. □

The first part of Equation (3) is a *first order quasilinear partial differential equation* on the $n$ components of the vector function $\rho_\pi$, and the second part is an additional algebraic constraint on these components. for all $u \in U$. We omit the subscript $\pi$ when clear from the context. Quasi-linear first-order PDE's appear in related questions in control theory, for instance [5] in Hamilton-Jacobi-Bellman's approach to optimal control and in feedback linearization based on Frobenius' Theorem. They may be in principle solved using the method of characteristics [34].

Our notion of equivariance generalizes a mathematical concept fundamental in group theory, and specifically in the symmetry analysis of nonlinear dynamical systems [35, 36] with the same name. A parametrized vector field $F$ is said to be $\rho$-equivariant, or $\rho$ is a symmetry of $F$, if, for each solution $z(t)$ of $\dot{z} = F(z, u)$, $\rho(z(t))$ is also a solution. This property is equivalent to the PDE $F(\rho(z), u) = \rho'(z) f(z, u)$ [35] that is part of the equivariance definition, when $\pi$ = identity. We are generalizing this concept in several ways. First of all, we must consider the far more general case $\pi \neq$ identity; in fact, in our context, $\pi$ = identity is not of any interest whatsoever, since we are precisely interested in the effect of the input transformations. Second, it is essential to our definition that we include the algebraic "boundary condition" $h(\rho_\pi(z)) = h(z)$, Finally, in dynamical systems one typically studies equivariances only with respect to Lie group actions. Moreover, since compact Lie groups acting on Euclidean space can be identified with subgroups of the orthogonal group $O(n)$, one finds in the classical definitions [35] only linear maps $\rho'(z) = \gamma z$ with $\gamma \in O(n)$ being considered (so $\rho'(z) = \gamma$).

## 2.4 Statement of main theorem

Our main result is as follows. It is proved in Section 4.

**Theorem 1** *An analytic and irreducible system is $\mathcal{P}$-invariant if and only if there exists a $\mathcal{P}$-equivariance family.*

An analytic system is one for which all the functions defining the system are real-analytic on the state coordinates. This means that they can be expanded into locally convergent power series around each point. Every function made up of elementary algebraic compositions of elementary functions is analytic; this includes all expressions involving polynomials and, more generally, any well-defined (no poles) rational functions (so mass-action kinetics and Hill-type models with any integer Hill coefficients are allowed) as well as trigonometric functions, exponentials, and logarithms in any combinations. All our examples are analytic, as long as we restrict expressions such as $u/x$ to $x > 0$ (or $x < 0$), so that there are no poles.

An irreducible system is one for which no conservation laws restrict motions to proper submanifolds (accessibility property) and no pairs of distinct states give rise to the same input/output behavior (observability property). We define precisely accessibility and observability in Section 4. Irreducibility is a weak technical assumption; we will show that these two properties hold for all the systems in Figs. 1 and 2. Irreducible systems, which are also called "minimal" or "canonical" in the control theory literature, are minimal, in the sense that no lower-dimensional subsystem has an identical input/output behavior [5, 37–39]. Analogous, but simpler, notions of irreducibility also appear in areas such as group representations (Schur's Lemma).

## 3   Examples of finding symmetries using the main theorem

We show here how Theorem 1 allows one to immediately determine invariance properties for large classes of two-dimensional systems, including the integral feedback and feedforward examples shown in Figs. 1 and 2. In all these cases, the PDE for equivariances, if there is one, can be easily solved for in closed form. We consider two-dimensional systems with output equal to one of the coordinates. We write $z = (x, y)$, $F(z, u) = (f(x, y, u), g(x, y, u))$:

$$\begin{aligned}
\dot{x} &= f(x, y, u) \\
\dot{y} &= g(x, y, u) \\
h(x, y) &= y
\end{aligned}$$

and wish to determine for which possible input set mappings $\pi : \mathbb{U} \to \mathbb{U}$ there is an associated equivariance $\rho = \rho_\pi$. We drop the subscript and write $\rho = (\rho^x, \rho^y)$. Since $h(x, y) = y$, the condition $h(\rho(x)) = h(x)$ says that $\rho^y(x, y) = y$. Thus finding $\rho$ is equivalent to finding its $x$-component, a function $\rho^x$ that satisfies:

$$\begin{aligned}
f(\rho^x(x, y), y, \pi u) &= \frac{\partial \rho^x}{\partial x}(x, y) f(x, y, u) \\
g(\rho^x(x, y), y, \pi u) &= g(x, y, u)
\end{aligned}$$

(no derivative in the second equation because, $\partial \rho^y / \partial y = 1$). This is a scalar first order quasi-linear PDE subject to a side algebraic "boundary condition".

We specialize next to special two cases that cover many examples of interest. Particular instances of the first case are the systems in Figs. 1(c,d) and 2(a). A particular instance of the second case is the system in Fig. 1(a). We assume that the systems in both of the next Lemmas are irreducible. For all our examples, irreducibility is checked in Section 4.

**Lemma 3.1**  Suppose that:

$$g(x, y, u) = G(u^\beta x^\mu, y)$$

and $G(\cdot, y)$ is one-to-one for each fixed $y$. (Assuming $x > 0$ if $\mu < 0$ or $u > 0$ if $\beta < 0$.) Then, the only possible symmetries are fold-changes $\pi u = pu$. Furthermore, the system is invariant under a set $\mathcal{P}$ of such symmetries if and only

$$p^{-\beta/\mu} f(x, y, u) \;=\; f(p^{-\beta/\mu} x, y, pu) \;\; \text{for all } x, y, u$$

and each $p$ in the set.

In the special case in which $\beta{=}1$ and $\mu{=}{-}1$, that is, if $g$ depends on the ratio $u/x$, this means that $f$ must satisfy:

$$p \, f(x, y, u) \;=\; f(px, y, pu)$$

and in the special case $\beta{=}\mu{=}1$, $f$ must satisfy $p^{-1} f(x, y, u) \;=\; f(p^{-1} x, y, pu)$. In either special case, if $f$ is independent of $u$, then response invariance to all scaling transformations ($\mathcal{P} = \mathbb{R}_{>0}$) is equivalent to the requirement that $f$ be homogeneous of degree 1 in $x$.

*Proof.* Since $G$ is one to one on $y$,

$$G\left((\pi u)^\beta (\rho^x(x, y))^\mu, y\right) \;=\; g(\rho^x(x, y), y, \pi u) \;=\; g(x, y, u) \;=\; G\left(u^\beta x^\mu, y\right) \;\; \text{for all } x, y, u$$

implies that:

$$(\pi u)^\beta (\rho^x(x, y))^\mu \;=\; u^\beta x^\mu \;\; \text{for all } x, y, u$$

or, equivalently:

$$\left(\frac{\pi u}{u}\right)^\beta \;=\; \left(\frac{\rho^x(x, y)}{x}\right)^{-\mu} \;\; \text{for all } x, y, u \,.$$

Define

$$p \;:=\; \left(\frac{\pi u_0}{u_0}\right)^{\frac{1}{\beta}}$$

for any fixed but arbitrary element $u_0 \in \mathbb{U}$. It follows that

$$\pi u = pu \;\; \text{and} \;\; \rho^x(x, y) = p^{-\beta/\mu} x \;\; \text{for all } x, y, u \,,$$

from which all the conclusions are immediate. ∎

**Lemma 3.2** Suppose that:

$$g(x, y, u) \;=\; G(\mu x + \beta u, y)$$

and $G(\cdot, y)$ is one-to-one for each fixed $y$. Then, the only possible symmetries are translations $\pi u = p + u$. Furthermore, the system is invariant under a set $\mathcal{P}$ of such symmetries if and only

$$f(x, y, u) \;=\; f(-(\beta/\mu)p + x, y, p + u) \;\; \text{for all } x, y, u \,.$$

and each $p$ in the set.

*Proof.* Since $G$ is one to one on $y$,

$$G\left(\mu\rho^x(x,y)+\beta\pi u, y\right) \;=\; g(\rho^x(x,y),y,\pi u) \;=\; g(x,y,u) \;=\; G\left(\mu x+\beta u, y\right)$$

implies that:

$$\mu\rho^x(x,y)+\beta\pi u \;=\; \mu x+\beta u \quad \text{for all } x,y,u$$

or, equivalently:

$$\pi u - u \;=\; -\frac{\mu}{\beta}(\rho^x(x,y)-x) \quad \text{for all } x,y,u\,.$$

Define

$$p \;:=\; \pi u_0 \,-\, u_0$$

for any fixed but arbitrary element $u_0 \in \mathbb{U}$. It follows that

$$\pi u = p + u \quad\text{and}\quad \rho^x(x,y) = -\frac{\beta p}{\mu} + x \quad \text{for all } x,y,u\,,$$

from which all the conclusions are immediate. ∎

We can now quickly classify the examples shown in Figs. 1 and 2.

The linear integral feedback system in Fig. 1(a) fits the form in Lemma 3.2, so it can only be $\mathcal{P}$-invariant with respect to transformations $u \mapsto p + u$, and the only possible equivariance is $\rho^x(x,y) = x + \beta p/\mu$. Since $f(x,y,u)$ is independent of $x$ and $u$, this is indeed an equivariance. This system is $\mathcal{P}$-invariant with respect to translations.

The systems in Fig. 1(c,d) and Fig. 2(a) all fit the form in Lemma 3.1, so they can only be $\mathcal{P}$-invariant with respect to scaling transformations $u \mapsto pu$, and the only invariance is equivalent to the condition

$$p^\varepsilon f(x,y,u) \;=\; f(p^\varepsilon x, y, pu)$$

where $\varepsilon$ is $+1$ and $-1$ for the systems in Fig. 1(c,d) respectively, and is $+1$ for the system in Fig. 2(a). In Fig. 1(c,d), the value of $\varepsilon$ is irrelevant, because $f(x,y,u)$ is independent of $u$ and is homogeneous of degree 1 in $x$, so the property holds. In Fig. 2(a), $f$ is homogeneous of degree 1 in $x$ and $u$ simultaneously, so again the property holds. In summary, all three systems are $\mathcal{P}$-invariant with respect to scalings $\mathcal{P}$.

The log-linear system in Fig. 1(b) is also $\mathcal{P}$-invariant for the set of scalings. This may be shown with the equivariance $\rho^x(x,y) = x + \beta \ln p/\mu$.

We remark that, generalizing Fig. 1(a) and Fig. 2(a), any $n$-dimensional linear system $\dot{x} = Ax + bu$ with a stable $A$ and $h(x) = cx$ such that $cA^{-1}b = 0$ (i.e., its DC gain is zero) is $\mathcal{P}$-invariant for $u \mapsto p + u$, with $\rho(x) = x - A^{-1}bp$. The corresponding log-linear system, in which $\dot{x} = Ax + b\ln u$, is invariant with respect to scalings.

Finally, we study the "sniffer" IFFL shown in Fig. 2(b). The equation "$g(\rho^x(x,y),y,\pi u) = g(x,y,u)$" means that $\beta\pi u - \gamma\rho^x(x,y)y = \beta u - \gamma xy$ for all $x,y,u$, and thus evaluating at $y = 0$ it follows that $\pi u = u$ (assuming $\beta \neq 0$). So no nontrivial $\mathcal{P}$-equivariance exists. By the necessity part of Theorem 1, we conclude that this system is *not $\mathcal{P}$-invariant for any possible $\mathcal{P}$*.

**Remark 3.3** Interestingly, although not invariant, the system in Fig. 2(b), as well as many other examples from [27], satisfy an "approximate" invariance property, in the following sense. Suppose that the $y$ variable varies faster than the $x$ variable, so that one may make a quasi-steady state approximation $\beta u - \gamma x y = 0$. This allows one to reduce to a one-dimensional system $\dot{x} = \alpha u - \delta x$ with output $y = (\beta/\gamma)(u/x)$. Scalings $u \mapsto pu$ and $x \mapsto px$ preserve the equations, thus showing that the reduced system is $\mathcal{P}$-invariant with respect to scaling. (More generally, given a linear system with an output that depends on the ratio $x_i/x_j$ of two variables $x_i$ and $x_j$, one obtains scale-invariance.) This means that the original system is "close" to having the scale invariance property. A precise statement can be made using singular perturbation theory. This observation was made several years ago in the context of models of *Dictyostelium* chemotaxis and neutrophils [40]. □

**Remark 3.4** An example with vector inputs is as follows. Suppose that we consider a vector integral feedback system of the following form:

$$
\begin{aligned}
\dot{x} &= (y - y_0)x \\
\dot{y} &= G(\langle u, x \rangle, y)
\end{aligned}
$$

with output $y$. The state-space $\mathbb{Z}$ is $\mathbb{R}^{n+1}$: $x$ and $u$ are $n$-dimensional real vectors and $y$ is scalar, and $\langle u, x \rangle$ denotes the inner (dot) product of $x$ and $u$. We claim that this system is $\mathcal{P}$-invariant with respect to the rotation group $\mathcal{P} = SO(n)$. Indeed, for each $R \in SO(n)$ we may define the equivariance $\rho_R(x, y)$ by mapping $(x, y)$ to the $n + 1$-vector with first $n$ components $Rx$ and last component $y$. Since $\rho_R$ is linear, and its partial derivative with respect to $y$ is 1 and its Jacobian with respect to the $x$ variables is $R$, we only need to check that $R(y - y_0)x = (y - y_0)Rx$, which is true because $y - y_0$ is a scalar, and that $G(\langle Ru, Rx \rangle, y) = G(\langle u, x \rangle, y)$, which is true because $R$ is an special orthogonal matrix. The exact same proof works for the larger orthogonal group (reflection/rotations) $O(n)$. We can also generalize to the case when $\mathcal{P}$ consists of completely arbitrary nonsingular transformations ($R \in GL(n)$). In that case, we would take $\rho_R(x) = (R^T)^{-1}x$ (inverse transpose) and use that $\langle Ru, (R^T)^{-1}x \rangle = \langle u, x \rangle$. □

# 4   Details and proof of main theorem

The main theorem is stated for systems that are irreducible, meaning that the both the accessibility and the observability properties must hold. We define precisely and discuss these two properties next.

## 4.1   The accessibility property and the accessibility Lie algebra

In order to define accessibility, we will need to employ the notion of accessibility Lie algebra associated to a system (1), which we briefly review here; see [5] for further details and basic properties. Recall first the notion of Lie bracket of two vector fields on a manifold, specialized here to open subsets of Euclidean space $\mathbb{R}^n$: for any two differentiable vector fields $f, g : \mathbb{Z} \to \mathbb{R}^n$ defined in an open set $\mathbb{Z} \subseteq \mathbb{R}^n$, their *Lie bracket* is the new vector field

$$[f, g] : \mathbb{Z} \to \mathbb{R}^n$$

defined by the formula

$$[f, g] := \frac{\partial g}{\partial x} f - \frac{\partial f}{\partial x} g.$$

When $f, g$ are twice differentiable, $[f, g]$ is again differentiable, and one can take further brackets such as $[[f, g], g]$. More generally, if the vector fields are smooth (that is, infinitely differentiable), one may take any number of iterated brackets.

For any subset of smooth vector fields $\mathcal{F}$ on $\mathbb{Z}$, the Lie algebra generated by $\mathcal{F}$, denoted as $\mathcal{F}_{LA}$, is defined as the intersection of all the Lie algebras of vector fields which contain $\mathcal{F}$. (The set of all such algebras is nonempty, since it includes the algebra of all vector fields on $\mathbb{Z}$.) Since the intersection of any family of Lie algebras is also a Lie algebra, it follows that $\mathcal{F}_{LA}$ is the smallest Lie algebra of vector fields which contains the set $\mathcal{F}$, and it coincides with the set obtained as follows. Denote $\mathcal{F}_0 := \mathcal{F}$, and, recursively, $\mathcal{F}_{k+1} := \{[f, g] \mid f \in \mathcal{F}_k, g \in \mathcal{F}\}$, $k = 0, 1, 2, \ldots$, as well as $\mathcal{F}_\infty := \bigcup_{k \geq 0} \mathcal{F}_k$. Then, $\mathcal{F}_{LA}$ is equal to the linear span of $\mathcal{F}_\infty$. (See [5], Lemma 4.1.4 for a proof of this equality.) In summary, every element in the Lie algebra generated by the set $\mathcal{F}$ can be expressed as a linear combination of iterated brackets of the form

$$[f_\ell, \ldots, [f_3, [f_2, f_1] \ldots]],$$

for some $f_i \in \mathcal{F}$.

If the system (1) is such that the state-space $\mathbb{Z}$ is an open subset of $\mathbb{R}^n$ and all the vector fields $\mathcal{F} = \{f(\cdot, u), u \in \mathbb{U}\}$ are smooth, we define its *accessibility Lie algebra* as $\mathcal{F}_{LA}$. For each $z_0 \in \mathbb{Z}$, one may consider the subspace $\mathcal{F}_{LA}(z_0) := \{X(z_0), X \in \mathcal{F}_{LA}\}$ of $\mathbb{R}^n$. The *accessibility rank condition* is said to hold for the system if

$$\mathcal{F}_{LA}(z_0) = \mathbb{R}^n \quad \text{for every } z_0 \in \mathbb{Z}.$$

For analytic systems, the accessibility condition is equivalent to the property that the set of points reachable from any given state $z$ has a nonempty interior, as follows by a result of Sussmann [37] that generalizes Nagano's theorem [41] on integrability of involutive families of vector fields; see a proof and more details in the textbook [5].

In the special case of input-affine systems, i.e. those defined by differential equations

$$\dot{x} = g_0(x) + u_1 g_1(x) + \ldots + u_m g_m(x) \tag{4}$$

where $g_i$, $i = 0, \ldots, m$ are $m + 1$ vector fields, it is not necessary to use all the vector fields in $\mathcal{F}$ when generating $\mathcal{F}_{LA}$, since $\mathcal{F}_{LA}(z_0) = \{g_0, \ldots, g_m\}_{LA}(z_0)$ for all $z_0$. See [5], Lemma 4.3.3 for a proof. (The technical condition in that Lemma that zero must belong to the input set $\mathbb{U}$ fails when inputs are asked to be strictly positive, as in many of our applications. However, the Lemma also holds when $\mathbb{U}$ consists of positive inputs, since a continuity argument makes clear that one obtains the same spans if 0 is added to $\mathbb{U}$.)

For example, we show next that all the systems in Figs. 1 and 2, which are all input-affine (with $m = 1$), satisfy the accessibility rank condition.

$\underline{\dot{x} = \alpha(y - y_0), \dot{y} = \beta u - \mu x - \gamma y.}$ Here:

$$g_0 = \begin{pmatrix} \alpha(y - y_0) \\ \mu x - \gamma y \end{pmatrix}, \qquad g_1 = \begin{pmatrix} 0 \\ \beta \end{pmatrix}, \qquad [g_0, g_1] = \begin{pmatrix} -\alpha\beta \\ \gamma\beta \end{pmatrix}.$$

Since the determinant of $(g_1, [g_0, g_1])$ equals $\alpha\beta^2 \neq 0$ at every $x$, the accessibility rank condition holds.

$\underline{\dot{x} = \alpha(y - y_0), \dot{y} = \beta \ln u - \mu x - \gamma y.}$ This is the same as the previous case, in so far as the accessibility condition is concerned, because the set of vector fields $\mathcal{F}$ is identical to the previous one.

$\underline{\dot{x} = \alpha x(y - y_0), \dot{y} = \beta \frac{u}{x} - \gamma y.}$ Here:

$$g_0 = \begin{pmatrix} \alpha x(y - y_0) \\ -\gamma y \end{pmatrix}, \qquad g_1 = \begin{pmatrix} 0 \\ \frac{\beta}{x} \end{pmatrix}, \qquad [g_0, g_1] = \begin{pmatrix} -\alpha\beta \\ -\frac{\alpha\beta}{x}(y - y_0) + \frac{\beta\gamma}{x} \end{pmatrix}.$$

Since the determinant of $(g_1, [g_0, g_1])$ equals $\alpha\beta^2/x \neq 0$ at every $x$, the accessibility rank condition holds.

$\underline{\dot{x} = \alpha x(y_0 - y), \dot{y} = \beta u x - \gamma y.}$ Here:

$$g_0 = \begin{pmatrix} \alpha x(y_0 - y) \\ -\gamma y \end{pmatrix}, \qquad g_1 = \begin{pmatrix} 0 \\ \beta x \end{pmatrix}, \qquad [g_0, g_1] = \begin{pmatrix} -\alpha\beta x^2 \\ \alpha\beta x(y - y_0) + \beta\gamma x \end{pmatrix}.$$

Since the determinant of $(g_1, [g_0, g_1])$ equals $\alpha\beta^2 x^3 \neq 0$ at every $x$, the accessibility rank condition holds.

$\underline{\dot{x} = \alpha u - \delta x, \dot{y} = \beta \frac{u}{x} - \gamma y.}$ Here:

$$g_0 = \begin{pmatrix} -\delta x \\ -\gamma y \end{pmatrix}, \qquad g_1 = \begin{pmatrix} \alpha \\ \frac{\beta}{x} \end{pmatrix}, \qquad [g_0, g_1] = \begin{pmatrix} \alpha\delta \\ \frac{\beta\delta}{x} + \frac{\beta\gamma}{x} \end{pmatrix}.$$

Since the determinant of $(g_1, [g_0, g_1])$ equals $\frac{\alpha\beta\gamma}{x} \neq 0$ at every $x$, the accessibility rank condition holds.

$\underline{\dot{x} = \alpha u - \delta x, \dot{y} = \beta u - \gamma x y.}$ Here:

$$g_0 = \begin{pmatrix} -\delta x \\ -\gamma x y \end{pmatrix}, \quad g_1 = \begin{pmatrix} \alpha \\ \beta \end{pmatrix}, \quad [g_0, g_1] = \begin{pmatrix} \alpha\delta \\ \alpha\gamma y + \beta\gamma x \end{pmatrix}, \quad [g_1, [g_0, g_1]] = \begin{pmatrix} 0 \\ 2\alpha\beta\gamma \end{pmatrix}.$$

Since the determinant of $(g_1, [g_1, [g_0, g_1]])$ equals $2\alpha^2\beta\gamma \neq 0$ at every $x$, the accessibility rank condition holds.

## 4.2 The observability property

A system (1) is said to be observable, or to have the *observability property*, if no two distinct states can give rise to an identical temporal response to all possible inputs. Formally:

$$\psi(t, z_0, u) = \psi(t, \widetilde{z_0}, u) \;\; \forall u, t \quad \Rightarrow \quad z_0 = \widetilde{z_0}.$$

For analytic input-affine systems (4) with output $h = (h_1, \ldots, h_p)$, one can restate the observability property as follows. The *observation space $O$* associated to the system is the vector space spanned by the set of all functions of the type:

$$L_{g_{i_1}} \ldots L_{g_{i_k}} h_j \tag{5}$$

(called the elementary observables of the system) over all possible sequences $i_1, \ldots, i_k$, $k \geq 0$, out of $\{0, \ldots, m\}$ and all $j = 1, \ldots, p$, where $L_X H = \nabla H \cdot X$ is the directional or Lie derivative of the function $H$ with respect to the vector field $X$ and one understands $L_Y L_X H$ as the iteration $L_Y(L_X H)$. We include the case in which $k = 0$, in which case the expression in (5) is simply $h_j$. Two states $z_1$ and $z_2$ are said to be *separated by $O$* if there exists some $H \in O$ such that $H(z_1) \neq H(z_2)$. Observability is equivalent to the property that any distinct two states can be separated by the observation space. See [5], Remark 6.4.2 for a proof and discussion.

For example, it is very easy to see that all the systems in Figs. 1 and 2, which are all input-affine (with $m = 1$ and $p = 1$), satisfy the observability condition. We must prove that if $H(z_1) = H(z_2)$ for every elementary observable (5), then $z_1 = z_2$. Since already with $k = 0$ we have that $y_1 = h(z_1) = h(z_2) = y_2$, it only remains to show that some linear combination of observables gives a one-to-one function of $x$. Note that $\nabla h = (0, 1)$, so the dot product $L_g H = \nabla H.g$ simply picks out the second coordinate of $H$.

$\underline{\dot{x} = \alpha(y - y_0), \dot{y} = \beta u - \mu x - \gamma y.}$ Here:

$$g_0 = \begin{pmatrix} \alpha(y - y_0) \\ \mu x - \gamma y \end{pmatrix}, \qquad g_1 = \begin{pmatrix} 0 \\ \beta \end{pmatrix}, \qquad L_{g_0} h = \mu x - \gamma y.$$

Since $L_{g_0} h + \gamma h = \mu x$ is one-to-one on $x$, the observability condition holds.

$\underline{\dot{x} = \alpha(y - y_0), \dot{y} = \beta \ln u - \mu x - \gamma y.}$ This is the same as the previous case, in so far as the observability condition is concerned, because $\psi(t, z_0, u)$ is the same as $\psi(t, z_0, \log u)$ for the previous system.

$\underline{\dot{x} = \alpha x(y - y_0), \dot{y} = \beta \frac{u}{x} - \gamma y.}$ Here:

$$g_0 = \begin{pmatrix} \alpha x(y - y_0) \\ -\gamma y \end{pmatrix}, \qquad g_1 = \begin{pmatrix} 0 \\ \frac{\beta}{x} \end{pmatrix}, \qquad L_{g_1} h = \frac{\beta}{x}.$$

Since $L_{g_1} h$ is one-to-one on $x$, the observability condition holds.

$\underline{\dot{x} = \alpha x(y_0 - y), \dot{y} = \beta u x - \gamma y.}$ Here:

$$g_0 = \begin{pmatrix} \alpha x(y - y_0) \\ -\gamma y \end{pmatrix}, \qquad g_1 = \begin{pmatrix} 0 \\ \beta x \end{pmatrix}, \qquad L_{g_1} h = \beta x.$$

Since $L_{g_1} h$ is one-to-one on $x$, the observability condition holds.

$\underline{\dot{x} = \alpha u - \delta x, \dot{y} = \beta \frac{u}{x} - \gamma y.}$ Here:

$$g_0 = \begin{pmatrix} -\delta x \\ -\gamma y \end{pmatrix}, \qquad g_1 = \begin{pmatrix} \alpha \\ \frac{\beta}{x} \end{pmatrix}, \qquad L_{g_1} h = \frac{\beta}{x}.$$

Since $L_{g_1} h$ is one-to-one on $x$, the observability condition holds.

$\underline{\dot{x} = \alpha u - \delta x, \dot{y} = \beta u - \gamma x y.}$ Here:

$$g_0 = \begin{pmatrix} -\delta x \\ -\gamma x y \end{pmatrix}, \qquad g_1 = \begin{pmatrix} \alpha \\ \beta \end{pmatrix}, \qquad L_{g_0} h = -\gamma x y, \qquad L_{g_1} L_{g_0} h = -\alpha \gamma y - \beta \gamma x.$$

Since $L_{g_1} L_{g_0} h + \alpha\gamma hy = -\beta\gamma x$ is one-to-one on $x$, the observability condition holds. (Observe that we cannot argue simply with $L_{g_0} h$, because the function $-\gamma xy$ is not one-to-one on $x$ in the special case $y = 0$.)

## 4.3   Proof of Theorem 1

We must prove that, for analytic and irreducible systems, existence of a $\mathcal{P}$-equivariance family is sufficient as well as necessary for $\mathcal{P}$-invariance.

### Sufficiency.

Suppose given a $\pi \in \mathcal{P}$ and an associated equivariance $\rho = \rho_p$. We claim that the steady-state mapping $\sigma$ interlaces $\pi$ and its associated $\rho = \rho_p$, in the sense that

$$\rho(\sigma(\bar{u})) = \sigma(\pi(\bar{u}))$$

for every $\bar{u} \in \mathbb{U}$. Indeed, from the property $F(\rho(z), \pi u) = \rho'(z)F(z, u)$, applied with any constant input $u(t) \equiv \bar{u}$, and $z = \sigma(\bar{u})$, it follows in particular that $F(\rho(\sigma(\bar{u})), \pi\bar{u}) = \rho'(\sigma(\bar{u}))F(\sigma(\bar{u}), \bar{u})$. Now, $F(\sigma(\bar{u}), \bar{u}) = 0$, by definition of $\sigma(\bar{u})$, so also $F(\rho(\sigma(\bar{u})), \pi\bar{u}) = 0$, which this means that $\rho(\sigma(\bar{u}))$ is the steady state $\sigma(\pi\bar{u})$ corresponding to the constant input $\pi\bar{u}$, as we claimed.

Now suppose that $z(t) = \varphi(t, \sigma(\bar{u}), u)$ solves $\dot{z} = F(z, u)$ with initial condition $z(0) = \sigma(\bar{u})$. Consider $z_*(t) = \rho_\pi(z(t))$. Computing the derivative $\dot{z}_*(t)$, and using the chain rule:

$$\dot{z}_*(t) = \rho_\pi'(z(t))\dot{z}(t) = \rho_\pi'(z(t))F(z(t), u(t)) = F(\rho_\pi(z(t)), \pi u(t)) = F(z_*(t), \pi u(t))$$

Moreover, $z_*(0) = \rho_\pi(\sigma(\bar{u})) = \sigma(\pi\bar{u})$, by the interlacing property. It follows that $z_*$ is the solution with initial condition correcponding to the "preadapted" value $\sigma(\pi\bar{u})$ and the input $\pi(u(t))$, i.e. $z_*(t) = \varphi(t, \sigma(\pi\bar{u}), \pi u)$. We conclude that

$$\psi(t, \sigma(\bar{u}), u) = h(z(t)) = h(\rho_\pi(z(t))) = \psi(t, \sigma(\pi\bar{u}), \pi u).$$

### Necessity.

Suppose given an analytic and irreducible system that is $\mathcal{P}$-invariant. Fix any $\pi \in \mathcal{P}$. We must find a differentiable mapping $\rho = \rho_\pi : \mathbb{Z} \to \mathbb{Z}$ such that (3) holds: $F(\rho(z), \pi u) = \rho'(z)F(z, u)$ and $h(\rho(z)) = h(z)$. Let us consider a modified system in which $G(z, u) = F(z, \pi u)$ and same output $k(z) = h(z)$. Then (3) asks that

$$\rho'(z)F(z, u) = G(\rho(z), u), \quad k(\rho(z)) = h(z) \qquad \forall z \in \mathbb{Z}, u \in \mathbb{U}. \tag{6}$$

In the language of [37], property (6) says that $\rho$ should be an isomorphism between the two systems:

$$\dot{z} = F(z, u), \quad y = h(z) \tag{7}$$

and

$$\dot{z} = G(z, u), \quad y = k(z) \tag{8}$$

or, equivalently, that the diagram in Fig. 5 should be commutative, where we write "$z \cdot u_t$" and "$z \cdot \pi u_t$" to denote the states $\phi(t, z, u)$ and $\phi(t, z, \pi u)$ respectively.

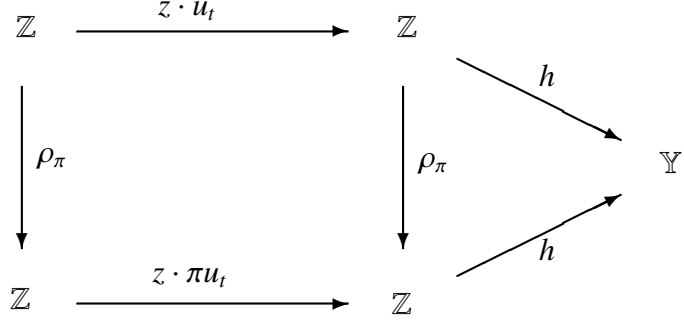

Figure 5: Equivariance as commutative diagram

Paraphrased in our language, Theorem 5 in [37] asserts the following. Suppose given any two analytic and irreducible systems (7) and (8), and two initial states $z_0$ and $\widetilde{z}_0$ respectively, with the property that the input/output behaviors are the same:

$$\psi(t, z_o, u) = \widetilde{\psi}(t, \widetilde{z}_o, u)$$

for all $t$ and all inputs $u$ (where we use tildes for the $\psi$ map of the second system), Then, there exists an isomorphism $\rho$ (that is, Equation (6) holds) such that also $\rho(z_o) = \widetilde{z}_0$. To apply this theorem to our case, we need to show that the modified system, with $G(z, u) = F(z, \pi u)$, is irreducible (it is clearly analytic, since the original system is) and we must find, for our original system and for $G(z, u) = F(z, \pi u)$, two respective initial states $z_0$ and $\widetilde{z}_0$ that lead to the same input/output behavior. This latter requirement is achieved by taking $z_0 = \sigma(u)$ for any fixed $u$, and $\widetilde{z}_0 = \sigma(\pi u)$ for the same $u$. The definition of $\mathcal{P}$-symmetry is precisely the statement that the two input/output behaviors are identical. We next verify that the modified system with $G(z, u) = F(z, \pi u)$ is irreducible.

*Accessibility:* Since $\pi$ is onto, the set of vector fields $\{F(\cdot, u), u \in \mathbb{U}\}$ is included in the set of vector fields $\{F(\cdot, \pi u), u \in \mathbb{U}\}$. Thus, the accessibility Lie algebra of the modified system can only be larger, and hence the accessibility property for the original system guarantees that of the modified system.

*Observability:* Suppose that two different states $z_1, z_2$ give rise to different outputs in the original system, when the input is $u(t)$. Since $\pi$ is onto, there is an input $v$ such that $u(t) = \pi(v(t))$ for all $t$. This means that the outputs corresponding to the initial states $z_1$ and $z_2$ in the modified system, under the input $v$, will be different. This argument can be applied for any two distinct states. Thus, the observability property for the original system guarantees that of the modified system.

This completes the proof of the main theorem. ∎

**Remark 4.1** The intuitive idea of the construction of $\rho$ is not difficult to understand, and is standard in control theory [5, 37–39]: given an initial state $x^0 = \sigma(\bar{u})$, input $u$, and time $t > 0$, we look at the state $z = \varphi(t, x^0, u)$ and the state $\widetilde{z} = \varphi(t, x^0, \pi u)$, and define $\rho(z) = \widetilde{z}$. The accessibility property says that this map is defined on an open subset of the state space. The interlacing property represented by Fig. 5 is checked by seeing that the states $\rho(z) \cdot u_t$ and $z \cdot \pi u_t$ are not distinguishable by any input/output experiments, and hence by observability must be the same. The technical difficulties arise in proving the differentiability of $\rho$ and its definition on the entire state space, not merely a subset, and this was one of the main contributions of [37]. □

**Remark 4.2** In the proof of the theorem, we only needed to know the existence of some two initial states $z_0$ and $\widetilde{z_0}$ that lead to identical behaviors. This means that, if we define a "weakly invariant" system as one for which there exists *some* constant $\bar{u}$ such that (2) holds: $\psi(t, \sigma(\bar{u}), u) = \psi(t, \sigma(\pi\bar{u}), \pi u)$ for all inputs $u$ and all $t \geq 0$ (instead of asking that this holds for every $\bar{u}$), then weak invariance implies the existence of an equivariance, and hence also invariance. (The irreducibility property plays a subtle role in this argument.) □

# 5   Stability result for nonlinear integral feedback systems

We wish to show the global asymptotic stability (GAS) of the unique steady-states $(\gamma\bar{y}/(\beta\mu), y_0)$ and $(\gamma\bar{y}/(\beta\mu), y_0)$ respectively, of the two nonlinear integral feedback systems in Fig. 1:

$$\begin{aligned} \dot{x} &= \alpha x(y - y_0) & \dot{x} &= \alpha x(y_0 - y) \\ \dot{y} &= \beta\frac{u}{x} - \gamma y & \dot{y} &= \beta u x - \gamma y \end{aligned}$$

for any constant input $u$, where we assume that $x(t) > 0$ and $u > 0$.

**Lemma 5.1** Consider any two-dimensional system evolving on $\mathbb{R}^2$ having the following "nonlinear damping" form:

$$\begin{aligned} \dot{x} &= g(y) \\ \dot{y} &= -f(x) - k(y)\,, \end{aligned}$$

where $f, g, k$ are functions that have positive derivatives. Suppose that $(x_0, y_0)$ is a steady state. Then this is the unique steady state of the system, and it is globally asymptotically stable.

**Corollary 5.2** Consider any two-dimensional system evolving on $\mathbb{R}_{>0} \times \mathbb{R}$ and having the following form:

$$\begin{aligned} \dot{x} &= x g(y) \\ \dot{y} &= -f(x) - k(y) \end{aligned}$$

where $k$ has positive derivatives and either (a) both $f$ and $g$ have positive derivatives or (b) both $f$ and $g$ have negative derivatives. Suppose that $(x_0, y_0)$ is a steady state. Then this is the unique steady state of the system, and it is globally asymptotically stable.

*Proof.* Suppose first that both $f$ and $g$ have positive derivatives. We transform the system into one in $\mathbb{R}^2$, using variables $y$ and $z = \ln x$:

$$\begin{aligned} \dot{z} &= \widetilde{g}(y) \\ \dot{y} &= -\widetilde{f}(z) - k(y) \end{aligned}$$

where $\widetilde{g}(y) = g(y)$ and $\widetilde{f}(z) := f(e^z)$. This functions defining this system have positive derivatives, as required in Lemma 5.1. Moreover, the transformed system has the steady state $(\ln x_0, y_0)$, which is therefore globally asymptotically stable (and unique). Transforming back to the original coordinates, we proved our claim. Now suppose that case (b) holds instead: both $f$ and $g$ have negative derivatives. We pick $z := -\ln x$. The equations transform as above, except that now $\widetilde{g}(y) = -g(y)$ and $\widetilde{f}(z) := f(e^{-z})$. Since these now have positive derivatives, the same argument as in case (a) applies. ∎

The nonlinear integral feedback systems discussed earlier are particular cases of the above form. The linear function $k(y) = \gamma y$ is increasing. When $g(y) = \alpha x(y - y_0)$, $g$ is increasing, and when $g(y) = \alpha x(y_0 - y)$, $g$ is decreasing. The function $f(x) = -\beta u/x$ (where $u$ is any positive constant) is increasing, while $f(x) = -\beta ux$ is decreasing.

*Proof of Lemma 5.1.* Uniqueness: since $g$ is strictly increasing, $y_0$ is uniquely determined, and it then follows that $x_0$ is also uniquely determined (because $f$ is increasing) as the solution of $f(x) = -k(y_0)$.

At the steady state $(x_0, y_0)$, we may assume, without loss of generality that $f(x_0) = 0$ and $k(y_0) = 0$ (as well as $g(y_0) = 0$). Indeed, let $c := f(x_0) = -k(y_0)$. Redefining $\widetilde{f}(x) := f(x) - c$ and $\widetilde{k}(y) := k(y) + c$, we have that $\widetilde{f}(x_0) = 0$ and $\widetilde{k}(y_0) = 0$, and the differential equations have not changed, since $-\widetilde{f}(x) - \widetilde{k}(y) = -(f(x) - c) - (k(y) + c) = -f(x) - k(y)$. Since $f$ is strictly increasing, its values are positive when $x > x_0$ and negative otherwise, and similarly for $g, k$ with respect to $y_0$. Let us define

$$V(x, y) := \int_{x_0}^{x} f(r)\, dr + \int_{y_0}^{y} g(r)\, dr\,.$$

By definition, $V(x_0, y_0) = 0$ and $V(x, y) > 0$ for all $(x, y) \neq (x_0, y_0)$ As $\frac{\partial^2 V}{\partial^2 x} = f'(x) > 0$, $\frac{\partial^2 V}{\partial^2 y} = g'(y) > 0$, and mixed second derivatives are zero, it follows that the Hessian matrix of $V$ is positive definite everywhere. Thus, $V$ is strictly convex, and it follows that $V$ is a proper function: $V(x, y) \to \infty$ as $\|(x, y)\| \to \infty$. In conclusion, $V$ is a Lyapunov function candidate function. To conclude global stability based on the LaSalle Invariance Principle [5], we must show that the derivative of $V$ along trajectories:

$$\dot{V}(x, y) := \frac{\partial V}{\partial x}(x, y) g(y) + \frac{\partial V}{\partial y}(x, y) \left[ -f(x) - k(y) \right]$$

has the properties that $\dot{V}(x, y) \leq 0$ for all $(x, y)$ and that $(x(t), y(t)) \equiv (x_0, y_0)$ is the only solution with $\dot{V}(x(t), y(t)) \equiv 0$. Now,

$$\dot{V}(x, y) = f(x) g(y) + g(y) \left[ -f(x) - k(y) \right] = -g(y) k(y) \leq 0$$

because $g$ and $k$ have everywhere the same sign (positive if $y > y_0$, negative if $y < y_0$). Suppose if a solution satisfies that $\dot{V}(x(t), y(t)) \equiv 0$, then $y(t) \equiv y_0$, so that also $\dot{y}(t) \equiv 0$, which substituted into the second equation gives $0 = -f(x(t)) - 0$, which implies that $x(t) \equiv x_0$. ∎

Observe that the only place that $k$ appears in the proof is in the statement that $g$ and $k$ have everywhere the same sign; thus the same proof works if instead of assuming that $k$ has everywhere positive derivatives, we assume that $f(x_0) = 0$ and $(y - y_0) k(y) > 0$ for all $y \neq y_0$. Note that $V$ would be a Hamiltonian for the system if $k \equiv 0$, so the system can be thought simply as adding damping to a conservative system.

# 6 Comparing feedforward and feedback structures

We make several remarks in this section concerning the relations and comparisons between adapting feedforward and feedback architectures.

## 6.1 Internal model principle

The "internal model principle" (IMP, for short) in control theory states that one should be able to recast any system which adapts to steps as a system which integrates an "adaptation error" signal (integral feedback). For example, it should be possible to rewrite the feedforward system in Fig. 2(a) in such a manner. In this section, we review one precise statement of the IMP, and apply it to this example.

Adaptation is called "disturbance rejection" in control theory [5] (not to be confused with a different topic, "adaptive control"). A key mathematical idea, the internal model principle (IMP), states that, to be able to adapt to all signals in a given class of inputs "$\mathcal{U}$", the system *must* include an "internal model": a subsystem which is driven by the "error" in adaptation, and whose solutions when the error is zero (that is, when the system has perfectly adapted) are the possible signals in $\mathcal{U}$. Intuitively, an internal representation of the external signal is memorized, and adaptation performance is constantly evaluated; any error in adaptation is used to form a better estimate of this external signal. For example, for adaptation to constant signals, the IMP requires integral feedback, as in Fig. 1: $\mathcal{U}$ = all constant signals, the error is $y - y_0$, and solutions of $\dot{x} = 0$ are precisely the constant signals. In systems biology, the IMP suggests biochemical structures, thus guiding modeling and experiments as well as interpretation of the role of various regulatory and signal processing motifs. For instance, the relevance of the IMP to *E. coli* chemotaxis was remarked in [42]: the methylation state can be viewed as a memory (integrator) and the "error" is the average kinase activity relative to its basal value. The IMP encompasses adaptation also with respect to richer classes of signals $\mathcal{U}$, not just constant ones. For example, one might speculate [43] that circadian rhythms might have evolved as an IMP mechanism to allow adaptation to day/night light and temperature cycles: a harmonic oscillator with period $T$ is predicted by the IMP when a system adapts to $\mathcal{U}$ = all signals of the form $A \sin(2\pi t/T + \varphi)$. The IMP was proved as a theorem for linear dynamics by Francis and Wonham in the mid 1970s [44,45]. It remains an open problem to find ultimate nonlinear generalizations, but there are some partial extensions known. For example, using "zero dynamics" ideas from [46], a theorem was given in [43] that shows, under appropriate technical assumptions, the existence of coordinate changes, generally nonlinear, that exhibit an internal model. Using this theorem, one should expect to find coordinate changes transforming IFFL circuits into integral feedback form, and this is indeed true. We first describe the comparatively trivial case of linear systems and then discuss how to obtain an analogous result for nonlinear IFFL systems.

As a simple first illustration, consider the following feedforward linear system:

$$\dot{x} = -x + u, \quad \dot{y} = -x - y + u,$$

which perfectly adapts to $y = y_0 = 0$ but is not in integral feedback form. We may perform a simple change of coordinates, representing the system using the state variables $(x = x - y, y)$ instead of the original $(x, y)$. In this new set of coordinates, we have:

$$\dot{x} = y, \quad \dot{y} = -x - 2y + u, \tag{9}$$

which is now an integral feedback system (the variable $z$ integrates the "error" $y$). We next review the main theorem from [43], and then work out the application to the nonlinear feedforward system in Fig. 2(a).

The general setup in [43] is as follows. The systems studied are scalar-input scalar-output $n$-dimensional systems for which the input appears to first order:

$$\dot{z} = f(z) + u g(z), \qquad y = h(z). \tag{10}$$

The vector fields $f$ and $g$ are smooth, and $h$ is a smooth function. We assume that $z = 0$ is a steady state when $u = 0$, $f(0, 0) = 0$.

We will say that the system (10) *adapts to inputs in a class* $\mathcal{U}$ if for each $u \in \mathcal{U}$ and each initial state $x^0 \in \mathbb{R}^n$, the solution of (10) with initial condition $x(0) = x^0$ exists for all $t \geq 0$ and is bounded, and the corresponding output $y(t) = h(x(t))$ converges to a fixed value $y_0 \in \mathbb{Y}$ (which does not depend on the particular input $u \in \mathcal{U}$) as $t \to \infty$.

In control theory, it is standard to describe the class of inputs $\mathcal{U}$ with respect to which adaptation holds through the specification of an "exosystem" that produces these inputs. An exosystem is simply any autonomous system $\Gamma$:

$$\dot{w} = Q(w), \quad u = \theta(w) \tag{11}$$

with the following property: the input class $\mathcal{U}$ consists exactly of the functions $u(t) = \theta(w(t))$, $t \geq 0$, for each possible initial condition $w(0)$. For example, if we are interested in step responses, we pick $\dot{w} = 0$, $u = w$. This means that the possible signals are the solutions of $\dot{w} = 0$, i.e. the constant functions of time; that is, $\mathcal{U}$ is the set of functions $u(t)$ for which $u(t) = \bar{u}$ for all $t$ for some $\bar{u} \in \mathbb{U}$. On the other hand, if we are interested in sinusoidals with frequency $\omega$ then we use $\dot{x}_1 = x_2$, $\dot{x}_2 = -\omega^2 x_1$, $u = x_1$. A technical assumption is that the signals in $\mathcal{U}$ do not grow without bound. Specifically, one assumes that the exosystem is *Poisson-stable*, meaning that for every state $w^0$, the solution $w(\cdot)$ of $\dot{w} = Q(w)$, $w(0) = w^0$ is defined for all $t > 0$ and it satisfies that $w^0$ is in the omega-limit set of $w$ (recall that this means that there is a sequence of times $t_i \to \infty$ such that the sequence $w(t_i)$ converges to $w^0$ as $t \to \infty$). In other words, the exosystem is almost-periodic in the sense that trajectories keep returning to neighborhoods of the initial state. Both the constant and sinusoidal examples mentioned above are generated by Poisson-stable systems. In contrast, ramps (linearly growing signals) are not generated by Poisson-stable systems, since they require an unstable second-order system $\dot{w}_1 = 0$, $\dot{w}_1 = w_2$, $u = w_1$ to generate them. Thus, the phenomenon of adaptation to ramps is not included in the scope of the theorem to be stated. The exosystem is assumed to have states that evolve on some differentiable manifold, $Q$ is a smooth vector field, and $\theta$ is a real-valued smooth function.

The IMP claims that a copy of this exosystem must be embedded in the system (10). More precisely, one says that the system *contains an output-driven internal model of* $\mathcal{U}$ if there is a change of coordinates which brings the equations (10) into the following block form:

$$\begin{aligned} \dot{z}_1 &= f_1(z_1, z_2) + u g_1(z_1, z_2) \\ \dot{z}_2 &= f_2(y, z_2) \\ y &= \kappa(z_1) \end{aligned} \tag{12}$$

so that the subsystem with state variables $z_2$ is capable of generating all the possible functions in $\mathcal{U}$: for some some function $\varphi(z_2)$, and for each possible $u \in \mathcal{U}$, there is some solution of

$$\dot{z}_2 = f_2(y_0, z_2) \tag{13}$$

which satisfies $\varphi(z_2(t)) \equiv u(t)$. "Change of coordinates" means that there is some integer $r \leq n$ and two differentiable manifolds $Z_1$ and $Z_2$ of dimensions $r$ and $n - r$ respectively, as well as a smooth function $\kappa : Z_1 \to \mathbb{R}$ and two vector fields $F$ and $G$ on $Z_1 \times Z_2$ which take the partitioned form

$$F = \begin{pmatrix} f_1(z_1, z_2) \\ f_2(\kappa(z_1), z_2) \end{pmatrix}, \quad G = \begin{pmatrix} g_1(z_1, z_2) \\ 0 \end{pmatrix}$$

and a diffeomorphism $\Phi : \mathbb{R}^n \to Z_1 \times Z_2$, such that

$$\Phi'(x)f(x) = F(\Phi(x)), \quad \Phi'(x)g(x) = G(\Phi(x)), \quad \kappa(\Phi_1(x)) = h(x)$$

for all $x \in \mathbb{R}^n$, where $\Phi_1$ is the $Z_1$-component of $\Phi$ and prime indicates Jacobian. Intuitively, the signal $z_2$ computes an integral of a function of the output $y(t)$, and when $y(t) \equiv y_0$, $z_2$ is (up to the mapping $\varphi$, which may be interpreted as a sort of rescaling) a signal in $\mathcal{U}$. For example, if $\mathcal{U}$ consists of constant functions (adaptation to steps), then for $y \equiv y_0$ one obtains (for different initial conditions) the possible constant signals.

In order to prove a theorem justifying the IMP, several technical conditions are imposed in [43]. The first is a signal detection or "sensitivity" property: (1) for some positive integer $r$, called in control theory a finite uniform relative degree,

$$L_g L_f^k h \equiv 0, \ k = 0, \ldots, r - 2 \qquad \text{and} \qquad L_g L_f^{r-1} h(x) \neq 0 \quad \forall x \in \mathbb{Z}.$$

As in the section on observability, generally, $L_X H$ denotes the directional or Lie derivative of a function $H$ along the direction of a vector field $X$: $(L_X H)(x) = \nabla H(x) \cdot X(x)$, and one understands $L_Y L_X H$ as the iteration $L_Y(L_X H)$. (In the special case that $L_g h(x) \neq 0$ for all $x$, the relative degree is $r = 1$, since the condition for $k < r - 1$ is vacuous.) Given that the relative degree is $r$, one may consider the following vector fields:

$$\widetilde{g}(x) = \frac{1}{L_g L_f^{r-1} h(x)} g(x), \quad \widetilde{f}(x) = f(x) - \left(L_f^r h(x)\right)\widetilde{g}(x), \quad \tau_i := \mathrm{ad}_{\widetilde{f}}^{i-1}\widetilde{g}, \ i = 1, \ldots r,$$

where $\mathrm{ad}_X Y$ is the operator $\mathrm{ad}_X Y = [X, Y] =$ Lie bracket of the vector fields $X$ and $Y$, and $\mathrm{ad}_{\widetilde{f}}^{i-1}$ is the iteration of this operator $i - 1$ times (when $i = 1$, $\tau_i = \widetilde{g}$). One says that a vector field $X$ is complete if the solution of the initial value problem $\dot{x} = X(x)$, $x(0) = x^0$ is defined for all $t$ and for any initial state $x^0$. Two vector fields $X$ and $Y$ are said to commute if $[X, Y] = 0$. The final assumptions, then, are that (2) $\tau_i$ is complete, for $i = 1, \ldots, r$ and (3) the vector fields $\tau_i$ commute with each other. (In the special case $r = 1$, condition (3) is automatic, since every vector field commutes with itself.) These assumptions are satisfied for linear systems. They are also satisfied, for example, for the feedforward system in Fig. 2(a):

$$\dot{x} = \alpha u - \delta x, \qquad \dot{y} = \beta \frac{u}{x} - \gamma y \tag{14}$$

with $h(x, y) = y$. In vector form, this is $\dot{z} = f(z) + u g(z)$, where the vector fields are:

$$f(x, y) = \begin{pmatrix} -\delta x \\ -\gamma y \end{pmatrix} \qquad \text{and} \qquad g(x, y) = \begin{pmatrix} \alpha \\ \beta/x \end{pmatrix}. \tag{15}$$

Since $L_g h = (0, 1) \cdot (\alpha, \beta/x)^{\mathsf{T}} = \beta/x$ is everywhere nonzero, we have that $r = 1$. Thus we only need to check that

$$\tau_1 = \widetilde{g} = \frac{1}{L_g h(x)} g(x) = \frac{x}{\beta} g(x) = \begin{pmatrix} \frac{\alpha}{\beta} x \\ 1 \end{pmatrix}$$

is complete, which is true because $\widetilde{g}$ is a linear vector field.

The main theorem in [43] says: *Suppose that assumptions (1)-(3) hold for the system (10). If (10) adapts to inputs in a class $\mathcal{U}$ generated by a Poisson-stable exosystem, then it contains an output-driven internal model of $\mathcal{U}$.*

The proof of the theorem consists of showing that there is, under the stated conditions, a change of variables as claimed. The map producing the change of variables is obtained by solving a first-order partial differential equation.

## 6.2 Illustration of IMP for the feedforward system in Fig. 2(a)

We consider the system (14), or (15) in vector form. We already checked properties (1)-(3), and the system adapts to steps (constant inputs), so the theorem says that it should be possible to to recast it integral feedback form. The proof in [43] asserts the existence of a mapping $\varphi(x, y)$ whose Lie-derivative along $g$ solves the following first-order linear PDE:

$$L_g \varphi = \nabla \varphi \cdot g = \alpha \varphi_x(x, y) + \frac{\beta}{x} \varphi_y(x, y) = 0.$$

Generally, such an equation may be solved using the method of characteristics. However, in our example the solution is immediate: $\varphi(x, y) = \alpha y - \beta \log x$. The map

$$(x, y) \mapsto (z_1, z_2) = (y, \varphi(x, y)) = (y, \alpha y - \beta \log x)$$

is a diffeomorphism whose inverse is $y = z_1$ and $x = e^{(\alpha z_1 - z_2)/\beta}$. We obtain the following equations in the new coordinates $(z_1, z_2)$:

$$\dot{z}_1 = \beta u e^{(z_2 - \alpha z_1)/\beta} - \gamma z_1$$
$$\dot{z}_2 = \beta \delta - \alpha \gamma z_1$$

with output $y = z_1$. This has the desired internal model form $\dot{z}_1 = f_1(z_1, z_2) + u g_1(z_1, z_2)$, $\dot{z}_2 = f_2(z_2, z_2)$, $y = \kappa(z_1)$, if we define: $f_1(z_1, z_2) = -\gamma z_1$, $g_1(z_1, z_2) = \beta e^{(z_2 - \alpha z_1)/\beta}$, $f_2(y, z_2) = f_2(y) = \beta \delta - \alpha \gamma y$, and $\kappa = $ identity. Thus $z_2$ is the variable that integrates the error: when $y = y_0 = 1$, the equation for $z_2$ becomes $z_2 = 0$, whose solutions are all the possible constant signals. We can also write this system in terms of the coordinates $x = e^{z_2/\beta}$, $y = z_1$ as follows:

$$\dot{x} = x \left( \delta - \frac{\alpha \gamma}{\beta} y \right), \qquad \dot{y} = \beta u x e^{-\frac{\alpha}{\beta} y} - \gamma y \tag{16}$$

which has the generic form $\dot{x} = x F(y)$, $\dot{y} = G(u x, y)$ of nonlinear integral feedback systems considered in Lemma 3.1.

### 6.3 What are the relative advantages of different architectures?

Both integral feedback and IFFL circuits allow adaptation, as well as, for appropriate models, symmetry invariance to scalings. Thus, it is natural to ask what are the relative merits of each of these architectures: what fitness-conferring signal processing and control properties are special for them? We view this as a question for further research, and limit ourselves here to a few remarks.

In a certain sense, the question is meaningless, since feedforward networks can often be simulated by feedback ones, as just shown. Nonetheless, the variable $\widetilde{x} = x - y$ may well be merely a mathematical construct with no biological meaning. In addition, any (linearized) system obtained by such a transformation from an IFFL is special: it can have *only real eigenvalues*, while the more general integral feedback form may have damped oscillatory behavior (depending on parameters).

This means that the feedback systems have a wider range of possible dynamical behaviors, and thus might be selected when it is desirable to meet specific performance objectives. In addition, feedback confers a certain robustness to uncertainty. We illustrate this point by comparing the feedforward system in Fig. 2(a):

$$\dot{x} = \alpha u - \delta x, \qquad \dot{y} = \beta \frac{u}{x} - \gamma y$$

with its recasting in feedback form:

$$\dot{x} = x\left(\delta - \frac{\alpha\gamma}{\beta}y\right), \qquad \dot{y} = \beta u x e^{-\frac{\alpha}{\beta}y} - \gamma y.$$

Both systems adapt ($y(t) \to \frac{\beta\delta}{\alpha\gamma}$). However, while in the feedback form, any perturbation of the right-hand side: $\dot{y} = \beta u x e^{-\frac{\alpha}{\beta}y} - \gamma y + \Delta(x, y)$ does not alter the property that the steady state must have $y = \frac{\beta\delta}{\alpha\gamma}$, no analogous simple statement can be made for the feedforward form.

Conversely, one may also speculate that biological or evolutionary could constrain the value of feedback structures. As an example, the stability of feedback (but not feedforward) systems is fragile to delays. Delays could arise from slower time scales for processes such as transcription and translation, compared to protein modifications. As an illustration, Fig. 6 show oscillations arising from a delay from $y$ to $x$ in the linear integral feedback system (9) and Fig. 7 show oscillations arising from a delay from $y$ to $x$ in the nonlinear integral feedback system (16).

## 7 Invariant steering

As remarked in [3], motile systems that measure a field in order to determine their velocity of movement have the property that their entire search patterns, as a function of time, are invariant to scale, if their sensory systems have the FCD property. We now discuss a precise formulation of this fact for arbitrary systems and symmetries.

We think of a system that, through its output $y(t)$, drives a steering mechanism ("motor complex"), resulting in a new position $r(t)$. We model this by a system with inputs, which is a way of saying that the position is computed by a dynamical system that keeps track of the past history of $y$:

$$\dot{q} = Q(q, y), \quad r = R(q)$$

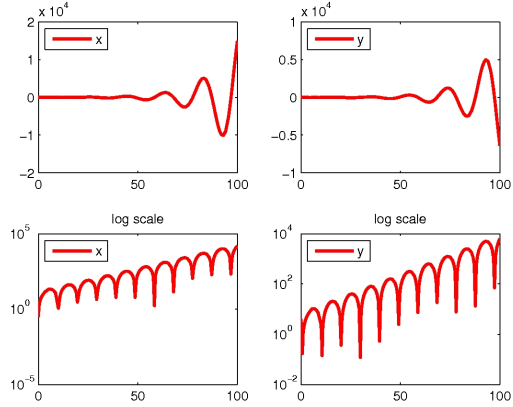

Figure 6: Oscillations in $\dot{x}(t) = y(t - h)$, $\dot{y}(t) = -x(t) - 2y(t) + u(t)$. Using $u \equiv 0$ and $h = 5$.

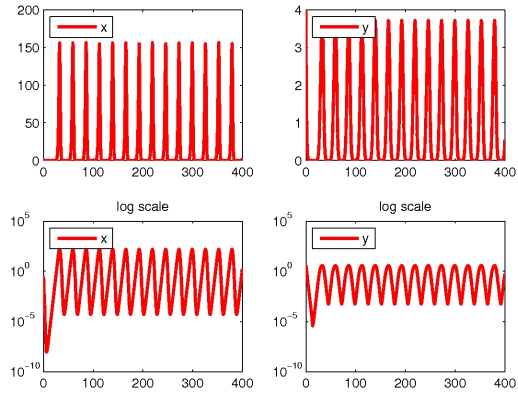

Figure 7: Oscillations in $\dot{x}(t) = x(1 - y(t - h))$, $\dot{y}(t) = x(t)u(t)e^{-y(t)} - y(t)$. Using $u \equiv 0$ and $h = 5$.

where $q(t)$ is the internal state of the steering mechanism. At position $r(t)$ in space, an "intensity" (e.g., light or nutrient concentration) is queried, and the result is a sensed input $I(t, r(t))$. The intensity field $I(t, r)$ could well be time as well as space-dependent. Finally, the loop is closed by the system measuring $I(t, r(t))$, except that we are interested in understanding how the system behaves if it measures $\pi I(t, r(t))$ instead of $I(t, r(t))$, where $\pi \in \mathcal{P}$ is a symmetry. See Fig. 8 for an illustration. We want to study the invariance of behavior of this system, and in particular of its position $r(t)$ as a function of time, under the assumption that the system had pre-adapted to a constant environment before $I(t, r) \equiv I_0$ when $t < 0$ before being placed in the current environment.

Formally, we start with a system that adapts and is invariant with respect to a set of symmetries $\mathcal{P}$, and consider the following extended system:

$$\begin{aligned} \dot{z} &= F(z, u) & u &= I(t, r) \\ \dot{q} &= Q(q, y) & y &= h(z), \ r = R(q) \,. \end{aligned}$$

We let $(z, q)$ be the solution with initial conditions $z(0) = \sigma(I_0)$ and $q(0) = q_0$, where $Q(q_0, y_0) = 0$, that is, $q_0$ is a steady state that corresponds to the adaptation value $y_0$ of the original system.

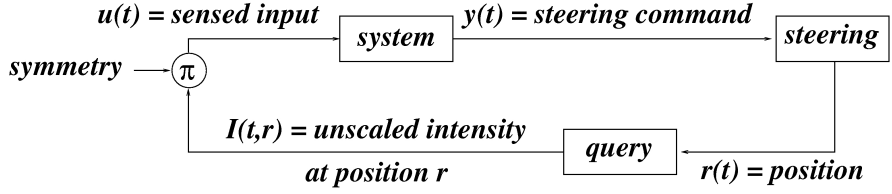

Figure 8: Closed-loop diagram for search under symmetry uncertainty for inputs

Also, for any given $\pi \in \mathcal{P}$, we consider the solution $(\widetilde{z}, \widetilde{q})$ of the system with initial conditions $z(0) = \sigma(\pi I_0)$ and $q(0) = q_0$ and intensity field $\pi I(t, r)$:

$$
\begin{aligned}
\dot{\widetilde{z}} &= F(z, \widetilde{u}) & \widetilde{u} &= \pi I(t, \widetilde{r}) \\
\dot{\widetilde{q}} &= Q(\widetilde{q}, \widetilde{y}) & \widetilde{y} &= h(\widetilde{z}), \ \widetilde{r} = R(\widetilde{q}) \, .
\end{aligned}
$$

We claim that $\widetilde{y}(t) = y(t)$, $\widetilde{u}(t) = \pi u(t)$, $\widetilde{r}(t) = r(t)$, and $\widetilde{q}(t) = q(t)$ for all $t \geq 0$.

To prove the claim, we consider the solution of the system

$$
\begin{aligned}
\dot{x} &= F(x, \pi u(t)) \\
\dot{s} &= Q(s, h(x))
\end{aligned}
$$

with initial conditions $x(0) = \sigma(\pi I_0)$ and $s(0) = q_0$. By definition of $\mathcal{P}$-invariance, we know that $h(x(t)) = h(z(t))$ for all $t \geq 0$. Thus, since the initial conditions on $q$ and $s$ are the same, it follows that also $s(t) = q(t)$ for all $t \geq 0$. Therefore, $u(t) = I(t, r(t)) = I(t, R(q(t))) = I(t, R(s(t)))$ for all $t \geq 0$. It follows that $\dot{x} = F(x, \pi I(t, r))$. We conclude that $(x, s)$ and $(\widetilde{z}, \widetilde{q})$ solve the same initial-value problem, and thus $x(t) = \widetilde{z}(t)$ and $s(t) = \widetilde{q}(t)$ for $t \geq 0$, from which $y(t) = h(z(t)) = h(x(t)) = h(\widetilde{z}(t)) = \widetilde{y}(t)$, and the claim is proved.

**Remark 7.1** A converse of this result holds as well. Suppose that, for every possible field $I$, the above system results in the same $r(t)$ when started from $z(0) = \sigma(I_0)$ as when starting from $z(0) = \sigma(\pi I_0)$ but using input $\pi I(t, r)$. This property then holds, in particular, when the field $I(t, r)$ is independent of position $r$, that is, $I$ is merely an arbitrary open-loop input. We would like to conclude that $y(t) = \widetilde{y}(t)$ from $r(t) = \rho(t)$, which means, since the input is arbitrary, that the original system must be invariant under $\pi$. This conclusion will hold provided that the steering system has the following property: if we solve $\dot{q} = Q(q, y)$ with initial condition $q(0) = q_0$ and two inputs $y_1$ and $y_2$, and the resulting solutions satisfy $R(q_1(t)) = R(q_2(T))$ for $t \geq 0$, then $y_1(t) = y_2(t)$ for $t \geq 0$. In control-theoretic terms, this input reconstruction property is stated as the requirement that the system $\dot{q} = Q(q, y)$ with output $R(q)$ be *input-observable*, which is a property closely related to right-invertibility, inverse dynamics, and "output to input observability" [46, 47]. Observability is almost enough to guarantee input-observability: if the system is observable, then $q_1(t) = q_2(t)$ for $t \geq 0$, so it is only needed that $Q(q(t), y_1(t)) = Q(q(t), y_2(t))$ imply $y_1(t) = y_2(t)$, which is a weak nondegeneracy property. □

**Remark 7.2** In many applications, the system output $y(t)$ drives a *stochastic* steering mechanism: the system producing the location $r(t)$ is subject to randomness, For example, in bacterial chemotaxis, $y(t)$ may represent a signal, such as the level of phosphorylated protein CheY, which serves to bias the random switches between tumbling and swimming behavior. One way to represent this probabilistic behavior is to model the dynamical system that computes the position from the history of $y(t)$ as:

$$\dot{q} = Q(q, y, X), \quad r = R(q)$$

where $X$ is a random process: $X = \{X_t(\omega), t \geq 0\}$ is defined on a probability space $(\Omega, \mathcal{F}, P)$, where $\Omega$ is a sample space, $\mathcal{F}$ is a $\sigma$-algebra of events, and $P$ is a probability measure; $P(X = v(t))$ is the probability of a given outcome (sample path) of this process. Under such a formalization, and assuming appropriate technical conditions, all the variables $(z(t), u(t), r(t), q(t), y(t))$ are themselves stochastic processes defined on the same probability space $\Omega$. (Technical conditions need to be imposed to insure the existence of solutions of the differential equations. We proceed intuitively, assuming that $X$ has piecewise continuous sample paths, thus avoiding complications of Itô calculus.) Now, given any fixed $\omega \in \Omega$, we may view $\dot{q} = Q(q, y, X)$ as a time-varying system $\dot{q}(t) = Q(q(t), y(t), t)$, where we have substituted $X = X_t(\omega)$ along this sample path. The previous proof extends, with no changes, to such a time-varying system. It follows that identical $r(t)$ (as well as $y(t)$ and $q(t)$) are obtained, whether using the field $I$ or using the changed field $\pi I$, so long as the initial condition $z(0)$ had also been modified by $\pi$. This holds for each $\omega \in \Omega$. Thus, as a random variable, $r(t)$ is invariant under $\mathcal{P}$. In particular, all statistics of $r$ remain invariant. □

## Acknowledgements

E.S. acknowledges support from grants from AFOSR FA9550-08, and NIH 1R01GM086881. Contributions: O.S, E.S. and U.A. participated in defining the questions, the examples, and their analysis; E.S. wrote the proofs.

# References

[1] L. Goentoro and M. W. Kirschner. Evidence that fold-change, and not absolute level, of $\beta$ -catenin dictates Wnt signaling. *Molecular Cell*, 36:872–884, 2009.

[2] C. Cohen-Saidon, A. A. Cohen, A. Sigal, Y. Liron, and U. Alon. Dynamics and variability of ERK2 response to EGF in individual living cells. *Molecular Cell*, pages 885–893, 2009.

[3] O. Shoval, L. Goentoro, Y. Hart, A. Mayo, E.D. Sontag, and U. Alon. Fold change detection and scalar symmetry of sensory input fields. *Proc Natl Acad Sci USA*, 107:15995–16000, 2010. Online before print doi: 10.1073/pnas.1002352107.

[4] L. Goentoro, O. Shoval, M. W. Kirschner, and U. Alon. The incoherent feedforward loop can provide fold-change detection in gene regulation. *Mol. Cell*, 36:894–899, 2009.

[5] E.D. Sontag. *Mathematical Control Theory. Deterministic Finite-Dimensional Systems*, volume 6 of *Texts in Applied Mathematics*. Springer-Verlag, New York, second edition, 1998.

[6] H. El-Samad, J. P. Goff, and M. Khammash. Calcium homeostasis and parturient hypocalcemia: An integral feedback perspective. *J. Theor. Biol.*, 214:17–29, 2002.

[7] P. Miller and X. J. Wang. Inhibitory control by an integral feedback signal in prefrontal cortex: A model of discrimination between sequential stimuli. *Proc. Natl. Acad. Sci.*, 103:201–206, 2006.

[8] K. V. Venkatesh, S. Bhartiya, and A. Ruhela. Mulitple feedback loops are key to a robust dynamic performance of tryptophan regulation in *Escherichia coli*. *FEBS Letters*, 563:234–240, 2004.

[9] T.-M. Yi, Y. Huang, M.I. Simon, and J. Doyle. Robust perfect adaptation in bacterial chemotaxis through integral feedback control. *Proc. Natl. Acad. Sci. U.S.A.*, 97:4649–4653, 2000.

[10] U. Alon. *An Introduction to Systems Biology: Design Principles of Biological Circuits*. Chapman & Hall, 2006.

[11] A. Ma'ayan, S. L. Jenkins, S. Neves, A. Hasseldine, E. Grace, B. Dubin-Thaler, N. J. Eungdamrong, G. Weng, P. T. Ram, J. J. Rice, A. Kershenbaum, G. A. Stolovitzky, R. D. Blitzer, and R. Iyengar. Formation of regulatory patterns during signal propagation in a Mammalian cellular network. *Science*, 309:1078–1083, Aug 2005.

[12] S. Mangan, S. Itzkovitz, A. Zaslaver, and U. Alon. The incoherent feed-forward loop accelerates the response-time of the gal system of Escherichia coli. *J. Mol. Biol.*, 356:1073–1081, Mar 2006.

[13] A. Cournac and J. A. Sepulchre. Simple molecular networks that respond optimally to timeperiodic stimulation. *BMC Syst Biol*, 3:29, 2009.

[14] G. Hornung and N. Barkai. Noise propagation and signaling sensitivity in biological networks: a role for positive feedback. *PLoS Comput. Biol.*, 4:e8, Jan 2008.

[15] S. Semsey, S. Krishna, K. Sneppen, and S. Adhya. Signal integration in the galactose network of Escherichia coli. *Mol. Microbiol.*, 65:465–476, Jul 2007.

[16] M. E. Wall, M. J. Dunlop, and W. S. Hlavacek. Multiple functions of a feed-forward-loop gene circuit. *J. Mol. Biol.*, 349:501–514, Jun 2005.

[17] D. Kim, Y. K. Kwon, and K. H. Cho. The biphasic behavior of incoherent feed-forward loops in biomolecular regulatory networks. *Bioessays*, 30:1204–1211, Nov 2008.

[18] S. Sasagawa, Y. Ozaki, K. Fujita, and S. Kuroda. Prediction and validation of the distinct dynamics of transient and sustained ERK activation. *Nat. Cell Biol.*, 7:365–373, Apr 2005.

[19] T. Nagashima, H. Shimodaira, K. Ide, T. Nakakuki, Y. Tani, K. Takahashi, N. Yumoto, and M. Hatakeyama. Quantitative transcriptional control of ErbB receptor signaling undergoes graded to biphasic response for cell differentiation. *J. Biol. Chem.*, 282:4045–4056, Feb 2007.

[20] P. Menè, G. Pugliese, F. Pricci, U. Di Mario, G. A. Cinotti, and F. Pugliese. High glucose level inhibits capacitative Ca2+ influx in cultured rat mesangial cells by a protein kinase C-dependent mechanism. *Diabetologia*, 40:521–527, May 1997.

[21] R. Nesher and E. Cerasi. Modeling phasic insulin release: immediate and time-dependent effects of glucose. *Diabetes*, 51 Suppl 1:S53–59, Feb 2002.

[22] M. P. Mahaut-Smith, S. J. Ennion, M. G. Rolf, and R. J. Evans. ADP is not an agonist at P2X(1) receptors: evidence for separate receptors stimulated by ATP and ADP on human platelets. *Br. J. Pharmacol.*, 131:108–114, Sep 2000.

[23] S. Marsigliante, M. G. Elia, B. Di Jeso, S. Greco, A. Muscella, and C. Storelli. Increase of [Ca(2+)](i) via activation of ATP receptors in PC-Cl3 rat thyroid cell line. *Cell. Signal.*, 14:61–67, Jan 2002.

[24] L. A. Ridnour, A. N. Windhausen, J. S. Isenberg, N. Yeung, D. D. Thomas, M. P. Vitek, D. D. Roberts, and D. A. Wink. Nitric oxide regulates matrix metalloproteinase-9 activity by guanylyl-cyclase-dependent and -independent pathways. *Proc. Natl. Acad. Sci. U.S.A.*, 104:16898–16903, Oct 2007.

[25] J. Tsang, J. Zhu, and A. van Oudenaarden. MicroRNA-mediated feedback and feedforward loops are recurrent network motifs in mammals. *Mol. Cell*, 26:753–767, Jun 2007.

[26] J.J. Tyson, K. Chen, and B. Novak. Sniffers, buzzers, toggles, and blinkers: dynamics of regulatory and signaling pathways in the cell. *Curr. Opin. Cell. Biol.*, 15:221–231, 2003.

[27] E.D. Sontag. Remarks on feedforward circuits, adaptation, and pulse memory. *IET Systems Biology*, 4:39–51, 2010.

[28] L. Yang and P.A. Iglesias. Positive feedback may cause the biphasic response observed in the chemoattractant-induced response of *dictyostelium* cells. *Systems Control Lett.*, 55(4):329–337, 2006.

[29] A. Levchenko and P.A. Iglesias. Models of eukaryotic gradient sensing: Application to chemotaxis of amoebae and neutrophils. *Biophys J.*, 82:50–63, 2002.

[30] F.-D. Xu, Z.-R. Liu, Z.-Y. Zhang, and J.-W. Shen. Robust and adaptive microRNA-mediated incoherent feedforward motifs. *Chinese Physics Letters*, 26(2):028701–3, February 2009.

[31] A. Kremling, K. Bettenbrock, and E. D. Gilles. A feed-forward loop guarantees robust behavior in escherichia coli carbohydrate uptake. *Bioinformatics*, 24:704–710, 2008.

[32] E. Voit, A. R. Neves, and H. Santos. The intricate side of systems biology. *Proc. Natl. Acad. Sci. U.S.A.*, 103:9452–9457, Jun 2006.

[33] L. Bleris, Z. Xie, D. Glass, A. Adadey, E.D. Sontag, and Y. Benenson. Synthetic incoherent feed-forward circuits show adaptation to the amount of their genetic template. 2010. Submitted.

[34] L. Evans. *Partial Differential Equations*. American Mathematical Society, Providence, 1998.

[35] M. Golubitsky and I. Stewart. *The Symmetry Perspective*. Birkhüser Verlag, Basel, 2002.

[36] J. Guckenheimer and P. Holmes. *Nonlinear Oscillations, Dynamical Systems, and Bifurcations of Vector Fields*. Springer-Verlag, New York, 1983.

[37] H.J. Sussmann. Existence and uniqueness of minimal realizations of nonlinear systems. *Math. Systems Theory*, 10:263–284, 1977.

[38] E.D. Sontag. *Polynomial Response Maps*, volume 13 of *Lecture Notes in Control and Information Sciences*. Springer-Verlag, Berlin, 1979.

[39] E.D. Sontag. Spaces of observables in nonlinear control. In *Proceedings of the International Congress of Mathematicians, Vol. 1, 2 (Zürich, 1994)*, pages 1532–1545, Basel, 1995. Birkhäuser.

[40] Pablo Iglesias, 2010. Personal communication.

[41] T. Nagano. Linear differential systems with singularities and an application to transitive lie algebras. *J. Math. Soc. Japan*, 18:398–404, 1966.

[42] T. M. Yi, Y. Huang, M. I. Simon, and J. Doyle. Robust perfect adaptation in bacterial chemotaxis through integral feedback control. *Proc. Natl. Acad. Sci. U.S.A.*, 97:4649–4653, 2000.

[43] E.D. Sontag. Adaptation and regulation with signal detection implies internal model. *Systems Control Lett.*, 50(2):119–126, 2003.

[44] B.A. Francis and W.M. Wonham. The internal model principle for linear multivariable regulators. *Appl. Math. Optim.*, 2:170–194, 1975.

[45] W.M. Wonham. *Linear Multivariable Control: A Geometric Approach, 3rd ed.* Springer-Verlag, New York, 1985.

[46] A. Isidori. *Nonlinear Control Systems*. Springer, London, 3rd edition, 1995.

[47] D. Liberzon, A. S. Morse, and E.D. Sontag. Output-input stability and minimum-phase non-linear systems. *IEEE Trans. Automat. Control*, 47(3):422–436, 2002.